\documentclass[sigconf]{acmart}
\AtBeginDocument{%
  \providecommand\BibTeX{{%
    \normalfont B\kern-0.5em{\scshape i\kern-0.25em b}\kern-0.8em\TeX}}}

\newcommand{\sysname}{\textsc{PaperWeaver}}
\newcommand{\sysws}{\textsc{PaperWeaver\ }}
\newcommand{\sys}{\textsc{PaperWeaver}}
\definecolor{interaction}{HTML}{2B8C40}

\newcommand{\eg}{\textit{e.g., }}
\newcommand{\cf}{\textit{cf. }}

\newcommand{\ie}{\textit{i.e., }}

\newcommand{\pmf}{\textit{contextualized aspect-based description}}
\newcommand{\rel}{\textit{paper-paper description}}

\definecolor{Issue1}{HTML}{08703D} 
\definecolor{Issue2}{HTML}{34217A} 
\definecolor{Issue3}{HTML}{BF1656} 
\definecolor{Issue4}{HTML}{4860cc} 

\newcommand{\rationale}[1]{#1}
\newcommand{\pipeline}[1]{#1}
\newcommand{\limitation}[1]{#1}
\newcommand{\discussion}[1]{#1}
\newcommand{\minor}[1]{#1}

\copyrightyear{2024}
\acmYear{2024}
\setcopyright{rightsretained}
\acmConference[CHI '24]{Proceedings of the CHI Conference on Human Factors in Computing Systems}{May 11--16, 2024}{Honolulu, HI, USA} \acmBooktitle{Proceedings of the CHI Conference on Human Factors in Computing Systems (CHI '24), May 11--16, 2024, Honolulu, HI, USA} 
\acmDOI{10.1145/3613904.3642196} 
\acmISBN{979-8-4007-0330-0/24/05}

\begin{document}

\title[\sysname: Enriching Paper Alerts by Contextualizing Descriptions]{\sysname: Enriching Topical Paper Alerts by Contextualizing Recommended Papers with User-collected Papers}

\author{Yoonjoo Lee}
\authornote{Work completed during a researcher internship at Semantic Scholar Research, Allen Institute for AI.}
\affiliation{%
  \institution{School of Computing, KAIST}
  \city{Daejeon}
  \country{Republic of Korea}
}
\email{yoonjoo.lee@kaist.ac.kr}

\author{Hyeonsu B. Kang}
\affiliation{%
  \institution{Human-Computer Interaction Institute, Carnegie Mellon University}
  \city{Pittsburgh}
  \state{PA}
  \country{USA}}
\email{hyeonsuk@andrew.cmu.edu}

\author{Matt Latzke}
\affiliation{%
  \institution{Allen Institute for AI}
  \city{Seattle}
  \state{WA}
  \country{USA}}
\email{mattl@allenai.org}

\author{Juho Kim}
\affiliation{%
  \institution{School of Computing, KAIST}
  \city{Daejeon}
  \country{Republic of Korea}}
\email{juhokim@kaist.ac.kr}

\author{Jonathan Bragg}
\affiliation{%
  \institution{Allen Institute for AI}
  \city{Seattle}
  \state{WA}
  \country{USA}}
\email{jbragg@allenai.org}

\author{Joseph Chee Chang}
\affiliation{%
  \institution{Allen Institute for AI}
  \city{Seattle}
  \state{WA}
  \country{USA}}
\email{josephc@allenai.org}

\author{Pao Siangliulue}
\affiliation{%
  \institution{Allen Institute for AI}
  \city{Seattle}
  \state{WA}
  \country{USA}}
\email{paos@allenai.org}

\renewcommand{\shortauthors}{Lee et al.}

\begin{abstract}
With the rapid growth of scholarly archives, researchers subscribe to ``paper alert’' systems that periodically provide them with recommendations of recently published papers that are similar to previously collected papers. However, researchers sometimes struggle to make sense of nuanced connections between recommended papers and their own research context, as existing systems only present paper titles and abstracts. To help researchers spot these connections, we present \sys{}, an enriched paper alerts system that provides contextualized text descriptions of recommended papers based on user-collected papers. \sys{} employs a computational method based on Large Language Models (LLMs) to infer users’ research interests from their collected papers, extract context-specific aspects of papers, and compare recommended and collected papers on these aspects. Our user study (N=15) showed that participants using \sys{} were able to better understand the relevance of recommended papers and triage them more confidently when compared to a baseline that presented the related work sections from recommended papers.

\end{abstract}





\maketitle

\section{Introduction}
\begin{figure*}[!h]
    \centering
    \includegraphics[width=\textwidth]{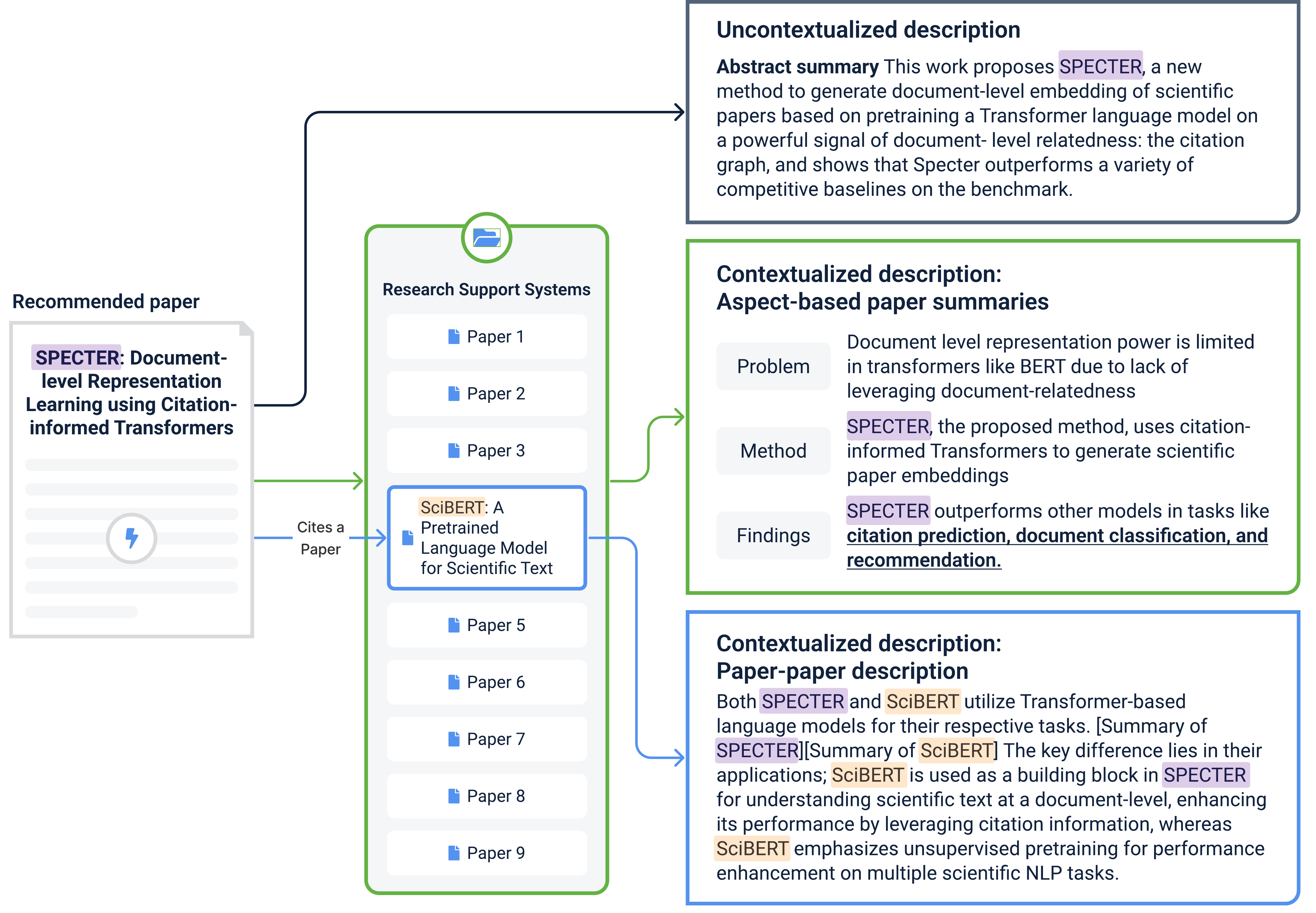}
    \caption{In contrast to existing topical paper alert systems that show descriptions with no personalized context for the recommended papers, \sys{} provides contextualized descriptions that surface the relevance of recommended papers and anchor them to familiar user-collected folders to help users better make sense of recommended papers.}
    \label{fig:teaser}
\end{figure*}

Managing an ever-increasing accumulation of knowledge has long been a challenge for scholars~\cite{blair2010too}. 
With the recent proliferation of published materials, researchers face an even bigger challenge of keeping up with the literature~\cite{Hemp2009DeathBI, Pain2016HowTK, Woolston2019PhDsTT}.
Fundamentally, scholars need to both discover new and relevant papers and contextualize them to their own research interests.
One popular practice recently is to leverage recommender systems that can help researchers retrieve potentially relevant, such as arXivist\footnote{\url{https://arxivist.com/}}, arXiv Sanity\footnote{\url{https://arxiv-sanity-lite.com/}} or Google Scholar\footnote{\url{https://scholar.google.com/}}. Typically, these systems allow users to create ``paper alerts'' by providing a short description of a specific research topic and a set of seed papers as examples of papers of interest.
For example, Semantic Scholar\footnote{\url{https://www.semanticscholar.org/}} allows users to save sets of collected papers under named topical folders. Users would then receive periodic paper alerts that contain a list of recently published papers similar to the collected papers.
\minor{These alerts help researchers to quickly narrow down from all recent publications to a small set of potentially relevant papers, allowing them to more easily stay up to date on research topics that are of interest to them.}

However, when receiving s set of potentially relevant papers in an alert, researchers still need to more deeply inspect each paper to understand its relevance.
This typically involves making meaningful connections 
between newly encountered information and their existing knowledge of the literature---a process that can incur high cognitive costs for researchers.
To illustrate this process with an example, a researcher working on \emph{Research Support Systems} may encounter
a relevant new paper titled ``SPECTER: Document-level Representation Learning using Citation-informed Transformers'' \cite{cohan-etal-2020-specter}. However, it can be challenging for the researcher to realize its relevance since the paper title only contains information about how it uses Transformers to generate document embeddings. Only when the researcher carefully examines the full abstract can they learn that the paper further described how the pre-trained model can be ``easily applied to develop downstream applications,'' and that the paper implemented a ``research paper recommender system'' for its evaluation. Further, if the researcher spent additional efforts to examine the content of the recommended paper, they might discover that the related work section described how the proposed method was built on SciBERT\cite{Beltagy2019SciBERTAP}---a familiar paper they had recently saved and read---but extends its capability from embedding sentences to long documents. \minor{This manual process is effortful but important because failure to identify meaningful connections between new and existing knowledge will lead to overlooking new papers relevant to a researcher's interests.}

Further, existing paper recommender systems typically only show a list of titles and abstract summaries for the recommended papers with little information on \emph{how} they were relevant to the topic of the folder or the set of seed papers that the recommendations were based on. Findings from our formative study (Section~\ref{sec:formativestudy}) suggest that the title of a paper often lacks enough details that help researchers understand the paper's relevance yet the abstract is often too long to skim through for paper alerts. Prior work has explored an approach that generates one-sentence TL;DR (too long; didn’t read) summaries \cite{Cachola2020TLDRES} that are easier to consume but they lack contextualization to the folder topic and the collected papers in them.  
Specifically, since the summaries are not tailored to a user's folder context, parts of the abstracts that showed how the recommended papers were relevant to the folder can sometimes be omitted.
While there has been research exploring ways to better contextualize paper recommendations by surfacing personalized social signals (e.g., based on a user's prior interaction or publication history~\cite{from_who_you_know}), they do not describe how the content of the papers is relevant to the users. As a result, users are left to decide whether to carefully examine the recommended papers to find potential connections, with no guarantee that the effort will pay off.
As shown in the above example, this sensemaking process can be effortful for users, and failure to effectively triage paper recommendations could result in overlooking important paper recommendations and reducing the effectiveness of such systems.

In this work, we investigate what types of information in paper alerts help scholars deeply understand the relevance of recommended papers to their topical folders.
In a formative study with seven researchers, we investigated challenges in identifying relevance from existing paper alerts and desired alternative descriptions of the recommended papers.
Our formative findings suggested that scholars desire paper alerts that were contextualized to their folders compared to only showing titles, abstracts, and uncontextualized summaries.
Participants found that descriptions about the recommended papers that revealed connections of a recommended paper to the user's folder context helped them understand the recommended paper more effectively since it spotlights where to focus on among many aspects of the paper. We also found that presenting the comparison and contrasting descriptions of multiple papers allowed participants to understand how the recommended papers build on prior work they had already collected in their folders. Anchoring unfamiliar papers with collected familiar papers also reduced the cognitive load of processing new information. 

\minor{Based on our formative findings,} we propose \sys, a new paper alert system that can enrich existing paper recommender systems by generating descriptions of how each recommended paper relates to a user's interests and their collection of papers. Our system is built on an existing document recommender system.
\sys\ leverages recent advancements in Large Language Models (LLMs) for text generation to support users in several ways.
First, \sys\ generates a compact topic description of a set of user-collected papers.
This compact topic description provides users with a quick summary of the collected papers. It is user-editable and useful for generating future descriptions for each recommended paper contextualized to a user's interest.
Second, to help users understand how a recommended paper is relevant to their research context, \sys\ generates two types of complementary contextual descriptions: \textit{contextualized aspect-based summaries} and \textit{paper-paper descriptions} (Fig.~\ref{fig:teaser}).
\textit{Contextualized aspect-based summaries} leverage the generated topic descriptions to extract statements on the problems, methods, and findings of papers that are highly relevant to the topic of interest.
\textit{Paper-paper descriptions} summarize how a recommended paper relates to collected papers, which are more familiar to the users. If the recommended paper cites collected papers, \sys\ summarizes these citation descriptions and, if not, it synthesizes relationships by comparing and contrasting aspects from the papers.
Motivated by how multiple alternative descriptions can improve understanding of complex scientific topics \cite{ainsworth2006deft,ainsworth2008educational}, we design an interactive paper alert interface where users can explore multiple descriptions for recommended papers (Fig.~\ref{fig:interface}).

To evaluate \sys{}, we conducted a within-subjects study ($N$ = 15) with researchers who were interested in receiving paper recommendation alerts. To ensure participants were motivated in the study, the paper alerts used in the study were generated based on their actual set of collected papers. We compared \sys{} with a strong baseline similar to existing paper alert systems but additionally enriched with uncontextualized summaries and extracted related work sections from the recommended papers. 
Our user study results showed that participants were able to better understand the nuanced relevance of the recommended papers and were able to triage them more confidently with \sys{}. Further, participants were able to capture richer relationships between recommended and collected papers in their notes when using \sys{} compared to the baseline.

The contributions of this work are as follows:


\begin{itemize}
    \item Qualitative findings from a formative study employing design probes with researchers that identified user challenges around making sense of recommended papers and the need for contextualized summaries.
    
    \minor{\item \sys{}, a tool that provides additional contextualized descriptions for a set of recommended papers, tailored to the user-collected papers. \sys{} uses an LLM-based pipeline that synthesizes content from both recommended and user-collected papers.}
    \item Findings from a user study ($N=15$) that demonstrated how using \sys{} facilitates sense-making of paper recommendations and aids in uncovering useful relationships between recommended and collected papers.
\end{itemize}

\section{Related Work}

Due to the increasing difficulties for scholars to keep up with the rapidly growing scholarly publications, significant research has been devoted to supporting scholars in better understanding the literature. Most prior work in scholar support tools has focused on two stages of this process: \textbf{broadening} scholars' reach in discovering relevant papers (\S\ref{subsection:related_work_breath}) and helping scholars to \textbf{deeply} understand the literature as they read (\S\ref{subsection:related_work_depth}). \sys{} focuses on a less explored area that bridges between the two stages: providing lightweight and contextualized sensemaking support when scholars are presented with a set of paper recommendation alerts. In this section, we briefly describe the two major threads of research to situate this current paper within the literature. 
\minor{Finally, we present prior work on contextualization in the settings where users acquire new knowledge and how they influence and differ from our approach (\S\ref{subsection:related_work_contextualization}).}

\subsection{Facilitating Broad Scholarly Exploration and Paper Discovery}
\label{subsection:related_work_breath}
Significant research has focused on helping scholars explore and discover relevant papers in the literature.
For example, SPECTER~\cite{cohan-etal-2020-specter,Singh2022SciRepEvalAM} leveraged citations between documents to train dense vector representations to encode the content-similarity of research papers. The vectors can then be used to power different downstream applications such as search or recommender systems.
Prior research \minor{has} also leveraged alternative similarity signals such as co-citations~\cite{paperquest}, domains~\cite{naacl2022_retrieval_across_domains}, and authorships~\cite{comlittee,bridger} to broaden the range of recommendations.
Recent work has begun to explore facilitating research paper retrieval with a deeper understanding of its semantic content. For example, \cite{chan2018solvent,kang_augmenting_tochi,hope_kdd17} focused on extracting a \emph{problem-method} schema from paper, which is rooted in cognitive science theories of analogical processing and creativity~\cite{gickholyoak1980,analogical_reminding}.
Diversifying retrieval based on one aspect (\eg mechanisms used to tackle a problem) while constraining the similarity on another (\eg the problems tackled in a paper), can effectively broaden recommendations across different domains and increase creativity in scientific ideation~\cite{kang_augmenting_tochi}.

At the same time, research has suggested that users can still be overwhelmed by the number of paper recommendations they receive~\cite{from_who_you_know} and pointed to the importance of helping user \emph{prioritize and understand} paper recommendations.
A research thread more closely related to our work has focused on enriching recommendations with lightweight relevance signals which can help scholars efficiently understand connections to other work and effectively triage them.
For example, prior work has examined presenting spatial overviews of paper collections based on semantic similarity (\cf Apolo~\cite{apolo}, LitSense~\cite{litsense}, FeedLens~\cite{feedlens}) and surfacing personalized social signals based on coauthorship and citations~\cite{from_who_you_know} or publication venues and institutions~\cite{feedlens}.

However, most previous systems either rely on external structured signals that are easy to understand but are divorced from the content of the papers (\eg ``you have cited the authors before'' \cite{from_who_you_know}) or latent semantic signals that are based on the content of the papers but difficult to understand (\eg clusters of papers based on embedding distances\footnote{\url{https://www.connectedpapers.com/}}). In contrast, \sysws draws from literature on schematic processing~\cite{gick_schema_1983} and leverages recent advancements in LLMs to extract problem-method-finding aspects across papers and generate easy-to-understand compare and contrasting statements that anchor the recommended papers to the user's familiar papers and context.
Furthermore, while in evaluation we used a specific existing implementation of paper recommender system, our proposed post-hoc enrichment technique can potentially be \minor{applied} to other paper discovery techniques outlined above. 

\subsection{Reusing Related Work Sections for Deeper Scholarly Sensemaking} \label{subsection:related_work_depth}
\label{subsection:related_work_depth}

There is a recent thread of research focusing on how to \emph{reuse} content from the related work sections of published papers to help scholars deeply understand the literature. For example, Relatedly~\cite{relatedly} allowed its users to search across related work sections extracted from many published papers and provided support for reading scattered paragraphs. It highlighted how rich synthesis contained in related work sections can help scholars better understand the landscape of a research field while gaining a deep understanding of how different prior work compare and contrast with one another. More closely related to this current work, CiteSee~\cite{chang2023citesee} and CiteRead~\cite{rachatasumrit2022citeread} are two paper reading tools that extracted citing sentences, or \emph{citances} as coined by~\cite{nakov2004citances}, from other papers to provide in-situ sensemaking support when reading a new paper. Particularly, CiteRead presented incoming citances as margin notes in relevant regions of the paper a user was reading to help them better contextualize the current paper with relevant follow-on work.
Another line of research focused on helping users clip citances and organize them into notable threads of research while reading papers to build up a better understanding of a field.
For example, Threddy~\cite{Kang2022ThreddyAI} supported dynamic clipping and organization while reading to help preserve users' context of reading when switching between different papers, and Synergi~\cite{synergi} extended this idea to automatically structure a thread-based hierarchy of relevant prior work, contextualized to user-selected citances in a paper.

\minor{In contrast to prior work that focused on reusing related work sections to support users when they are deeply engaged in reading and reviewing the literature, this paper explores how to support users by enriching paper alerts to better understand and triage recommended papers.} The opportunity we exploit here is that recommended papers in a paper alert are likely to cite and describe how they compare and contrast with the user-collected papers in their related work sections. Specifically, related work sections typically start with contextualized summaries of relevant prior work (which could contain one or more seed papers) and end with a contrasting statement to differentiate their work.
As an example to illustrate this opportunity, in this current section, we first summarized how CiteRead (i.e., a user-collected paper) also reused contents of existing related work sections and ended with a contrasting statement that it focused on supporting deep reading as opposed to this current paper (i.e., a recommended paper) which focuses on enriching paper recommendation alerts. \sys{} leverages LLMs to extract this relationship from related work sections of recommended papers to generate a short contrasting statement so that users can easily anchor unfamiliar papers to familiar contexts, allowing them to see meaningful connections between papers in the literature.

\subsection{Providing contextualized explanation}\label{subsection:related_work_contextualization}
Knowledge acquisition requires assimilating new information into existing knowledge~\cite{ausubel2012acquisition}. Existing work across different domains has explored mechanisms and effects of explanations that are contextualized to a user's existing knowledge.
Prior research has considered various definitions of \textit{context}.
For example, Tutorons~\cite{head2015tutorons} and ScholarPhi~\cite{head2021augmenting} provide explanations of code and scientific notation, respectively, that are relevant to the user's reading context.
To explain scientific concepts to children, DAPIE~\cite{lee2023dapie} adapts how it explains concepts (\eg through simplification or examples) according to a child's responses in previous dialogue turns.
To provide adaptive explanations even when contextual information about the user is unavailable, AXIS~\cite{williams2016axis} leverages other users' ratings on provided explanations to identify effective explanations that can be provided to future users.
Recent work in NLP~\cite{salemi2023lamp} has investigated how to guide LLMs to personalize their generated text based on ``user profiles'’ (\eg previous data or content that a user has written) which can help to explain, paraphrase, or summarize text into language that is familiar to the user. 
Building on this thread of research, our work investigates an approach to synthesizing papers collected by a scholar into folder descriptions. These descriptions are then leveraged as representations of both the user's interest and prior understanding when generating descriptions to contextually explain new paper recommendations.
Regarding scientific literature understanding, ACCoRD\cite{Murthy2022ACCoRDAM} defines unfamiliar scientific concepts in terms of different reference concepts by taking advantage of diverse ways a concept is mentioned across the scientific literature. Notably, this work found that users prefer multiple descriptions to a single best description.
At a high level, ACCoRD and \sys{} both aim to make unfamiliar ideas more accessible by creating bridges upon the user's prior knowledge. However, the approach explored in ACCoRD operates at the unit of concepts, while \sys{} aims to help scholars understand the relevance between different papers that may employ many interconnected concepts and other aspects (\eg problem, method, findings) of academic papers.

\section{Formative Study}\label{sec:formativestudy}
We conducted a formative study with two goals. Firstly, to learn about researchers' current challenges when reading existing paper alerts that consist of a list of titles and abstract. Secondly, to gain \minor{a} deeper understanding of the types of information that would be more useful in a paper alert setting to inform the design of \sys{}.


\subsection{Procedure and Apparatus}

We conducted a targeted recruitment of seven participants who are graduate students expressed interest or have worked on one of two predefined topics: ``Reading support tools for scientific articles'' or ``Scholarly document processing.'' Four participants identified their discipline as human-computer interaction (HCI) and three as natural language processing (NLP).\footnote{Two participants joined in person and the rest joined remotely through Google Meet. This study was approved by our internal review board. Each participant was paid 30 USD for one hour of their time.}
To seed the two folders with papers that \minor{the participants are likely to be familiar with}, we chose relevant papers published by their research groups. To generate paper recommendations for each of the two topics, we used a publicly available paper recommender API with the seed papers as inputs.\footnote{\url{https://api.semanticscholar.org/api-docs/recommendations}} 

We designed five different types of contextualized text descriptions as design probes to gather feedback. The design probes covered different ways of describing the recommended papers in the context of a paper alert. The descriptions were generated either manually or using a previous scientific abstract summarization method called TL;DR~\cite{Cachola2020TLDRES} and an LLM with prompts that focused on including different types of information, such as relevance to the folder via the folder name or relevance to the seed papers (List of design probes in Supplementary Materials).

To simulate realistic paper alert usage, we asked participants to consider the pre-defined folders as their own. In the first 15 minutes of the interview, participants reviewed 10 seed papers on the topic they had chosen to ensure that they are familiar with the papers we had collected for them. We then probed their experience with current paper alert systems by showing them recommended papers in the standard presentation as a list of titles and abstracts, and asked them about their goals and pain points.
Afterward, we conducted a think-aloud interview by walking them through a set of design mock-ups with the design probes for 40 minutes. We probed the costs and benefits based on their reactions to different types of descriptions. Participants were also asked to participate in co-design by revising the descriptions as they saw fit. The interviews were screen recorded, and the first author conducted a thematic analysis to capture qualitative insights \cite{Boyatzis1998TransformingQI,Connelly2013GroundedT}.
The rest of this section lists the common themes from the interviews and describes the design probes when relevant.

\subsection{Formative Study Findings}


\subsubsection{Challenges in Making Sense of Existing Paper Alerts}
 
When we showed a design probe similar to existing paper alert systems \minor{with} a list of titles and abstracts for the recommended papers, all participants mentioned that the biggest challenge is to figure out how all the recommended papers are relevant to the topic of their folders\minor{. It is hard to quickly decide} whether they should save and read the recommended papers \minor{from just their titles and abstracts}. In order to effectively identify relevance, participants pointed to how they \minor{need} to carefully read the full abstracts or even the papers themselves. However, \minor{doing so} would take more time and effort than they typically spend on consuming paper recommendation alerts. The cost of carefully reading a list of full abstracts is so high that many participants resort to using \minor{a less effective} strategy where they spot mentions of certain keywords in the titles or whether it was published by a familiar author (P1, 3, 4, and 6). At the same time, they acknowledge that this strategy leads to overlooking relevant papers. Participants pointed to two reasons: (1) They might not know all relevant authors in the space which suggests previous social-signal-based approaches could be insufficient \cite{from_who_you_know}. (2) The keyword spotting only allows them to find high-level topical relevance (e.g., ``scientific papers'') but does not allow them to find deeper connections hidden in the abstracts (e.g., ``semantic embeddings that can support citation predictions'' \cite{cohan-etal-2020-specter}). 

\subsubsection{Surfacing Relevant Aspects when Summarizing Recommended Papers}
We explicitly asked participants to compare different types of description in three \minor{mock-ups,} each with different descriptions to the same set of paper recommendations: full abstracts, ``TL;DR'' abstract summaries generated using \cite{Cachola2020TLDRES}, and abstract summaries that focused on including information relevant to the folder name generated using a\minor{n} LLM. When comparing with the full abstracts, participants reacted positively with the two shorter abstract summaries over reading the full abstracts that were longer and less concise.
Additionally, they preferred the LLM-generated contextualized summaries over the uncontextualized TL;DR summaries. Specifically, participants felt that they could more effectively understand why the recommended papers were relevant while the TL;DR summaries often omitted parts of the abstract that were relevant to the folder. For example, P2 and P5 liked the contextualized sentence: ``\emph{You may be interested in this paper because it addresses the issue of trustworthiness and factuality in question answering systems, which can be relevant to processing scientific documents}[i.e., the folder name],'' which highlights parts of the abstract most relevant to the folder. Additionally, we observed that participants less familiar with the topic reacted particularly strongly with this type of explanation while \minor{more experienced} participants felt the descriptions were accurate but \minor{sometimes} trivial because they had more background knowledge in the domain.
\newpage{}

\subsubsection{Comparing and Contrasting Recommended and User-collected Papers}
Participants emphasized that the main goal when making sense of paper recommendations is to situate recently published papers within their own research context. 
With this goal, participants mentioned that the design probe with descriptions that focused on showing connections between recommended and collected papers helped them recognize recommended papers that directly \minor{build} on a paper they had collected. Additionally, anchoring recommended papers to a familiar collected paper helped them more effectively understand how they were situated within the literature. For example, P3 found a description that connected two papers by a similar research problem to be useful, and said ``By aligning the [similar] motivations of Foundwright [i.e., \cite{Park2023FoundWrightAS}; a recommended paper] with the motivation of Citesee[i.e., \cite{chang2023citesee}; a collected paper], I don't need to find and read two [separate] sentences that \minor{explain} the problem in the [two] abstracts.''

\subsubsection{Bringing New Perspectives to Previously Collected Papers}
One interesting observation was that participants found value not only in seeing information about the recommended papers, but their connections to papers in the folders also provided new insights about them (P1, P3, P6). For example, when seeing a design mock that showed the citation context (i.e., citing sentences) in a recently published recommended paper that cited one of the papers in the folder, participants felt that they had ``rediscovered'' the collected papers and how the descriptions ``makes me feel like the collected papers are still relevant.'' and that they ``[re]introduced papers that I already saved but had not read deeply'' providing them with ``new opportunities to better understand them and learn new perspectives about them'' (P1).

\subsection{Design Goals}
Based on insights from the formative interviews, we formalize the following Design Goals for our system:
\begin{itemize}
\item[\textbf{[DG1]}] Describe details about the recommended papers in a way that helps users understand how they are relevant to the topic of their \minor{research context (e.g., folders) } to avoid overlooking relevant recommended papers.
\item[\textbf{[DG2]}] Help users make connections between the recommended and collected papers by comparing and contrasting them.
\item[\textbf{[DG3]}] Reveal new aspects of previously collected papers to keep their understanding of the papers up-to-date and remind users of unread collected papers that have become more relevant with new surfaced connections.
\end{itemize}

\section{\sys{}}
\begin{figure*}[!h]
    \centering
    \includegraphics[width=\textwidth]{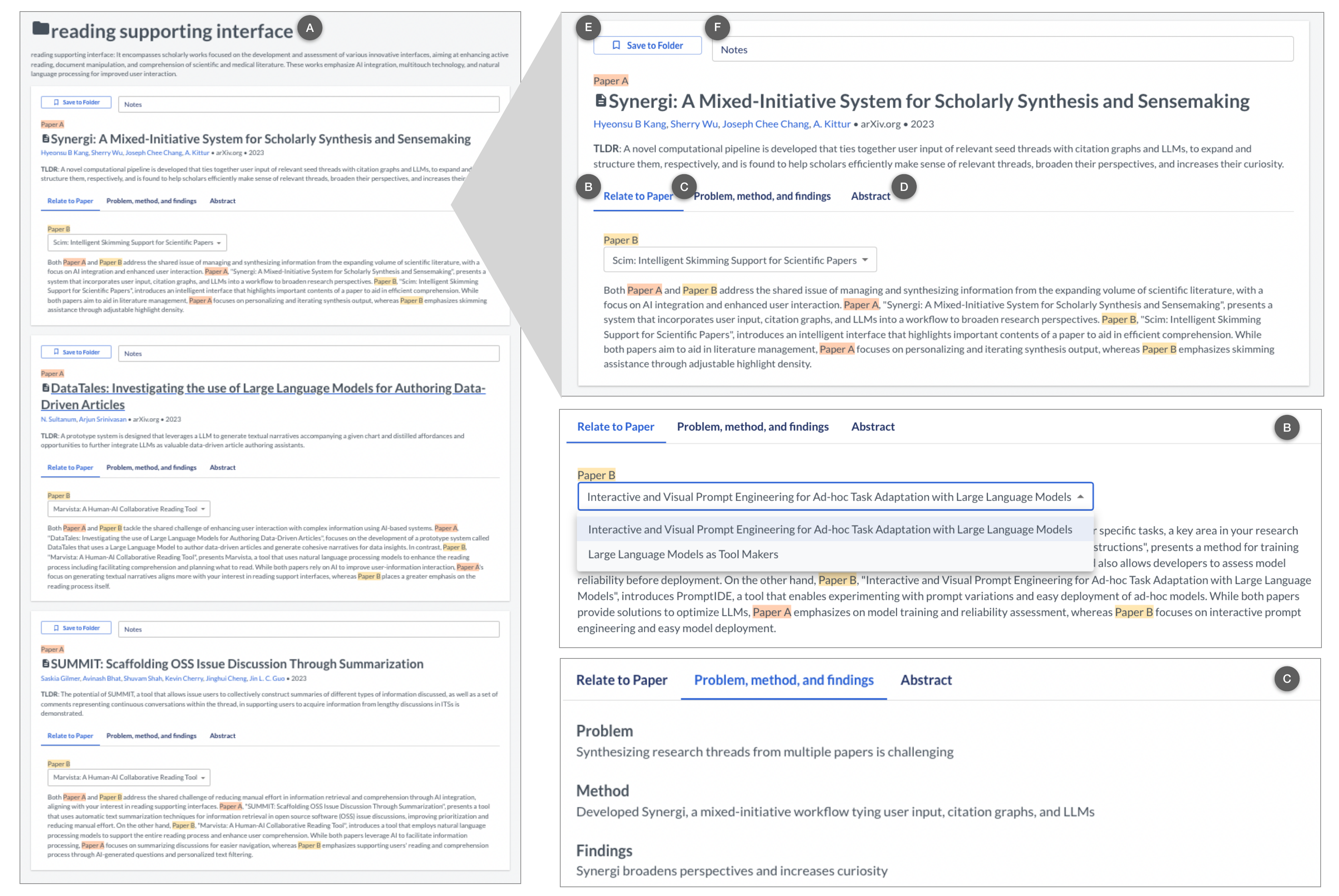}
    \caption{\sys{}'s paper alert interface. The page shows the folder title and description (A) and a list of recommended papers. Each recommended paper is shown on a paper card with the title, authors, venue, and publication year. The bottom of the card features description\minor{s} in three tabs: \textit{Relate to Paper} (B) allows users to select a paper from existing papers in the folder with the description related to the recommended paper. These descriptions are \textit{paper-paper Descriptions Based on Citances} and \textit{Paper-paper Descriptions via Generated Pseudo-citances} that focus on specific relationships between two papers. The description assigns the label "Paper A" to the recommended paper and "Paper B" to the existing paper in the library and highlights the references in different colors; \textit{Problem, method, and finding\minor{s}} (C) describes the recommended papers in three aspects--- problem, method, and findings---related to the folder context with \textit{contextualized Aspect-based Paper Summaries}; \textit{Abstract} (D) shows the unmodified abstract of the paper. The user can save a paper to the library (E) and take notes about the recommended paper (F).}
    \label{fig:interface}
\end{figure*}
Based on our design goals, we present \sys{} (Fig.~\ref{fig:interface}), a paper alert system that contextualizes recommended papers' descriptions based on a user's topical folder information. \sys{} enables users to explore these descriptions while reading paper alerts. \minor{With lessons learned from the formative study}, our computational method leverages a combination of a\minor{n} LLM\minor{-generated text} and extracted related work sections from the recommended papers. \sys{} extracts folder-specific aspects from \minor{a} recommended \minor{paper} to surface how it is relevant to \minor{the} user's context (DG1). 
The system describes how \minor{the} recommended paper is similar to and different from \minor{a} topical folder paper by identifying the relationships between papers (DG2). 
To remind users about papers they have previously collected in the folder, \sys{} includes information about both the recommended and the collected papers in the descriptions (DG3). 
Given a paper recommendation and a set of collected papers in a named topical folder, the system provides ``\textit{\minor{paper-paper} descriptions}'' that compare and contrast the recommended paper anchored to a relevant collected paper. To generate this \minor{type of descriptions}, the system use\minor{s} a pair of \minor{an} abstract and \minor{a} citance as input when available. For recommended papers that \minor{do} not cite any of the collected papers, \sys{} use\minor{s} \minor{an} LLM to identify latent relationships from their abstracts. Finally, \sys{} also generates ``\textit{\minor{contextualized aspect-based paper descriptions}}'' that summarizes a recommended paper's abstract in a way that reflects \minor{the paper's} relevance to the topic of \minor{the given} folder.

\subsection{Example User Scenario}

Imagine a computer science researcher who started working on the topic of \emph{Reading-supporting interfaces} a few weeks ago.
She has collected a list of papers related to this topic in a folder. Her familiarity with each paper in the folder varies. She has read some of the papers during her prior project. For some other papers, she only saved them because she saw relevant keyword\minor{s} in the title of the paper but has not had the time to read them yet. Because this is her active project, she is looking out for new papers relevant to the topic. She signed up for a paper alert service \minor{that} regularly provides her with a list of paper\minor{s} based on papers that she has collected. However, she felt that the information provided was lacking. She often had to wade into a paper’s abstract and its full text to see whether \minor{the paper} was relevant to her research interest\minor{s}. This activity requires more time and attention than she usually has when reading paper alerts.

Looking for a more effective way to process paper alerts, she gives \sys{} a try. She starts using the system by creating a folder titled \emph{Reading-supporting interfaces} and adding the list of papers she has collected to the folder. \sys{} automatically expanded on the folder title to provide an overview summary of the source papers she collected. Although she is mostly satisfied with the automated summary, she removes the keyword \emph{medical literature} in the description and adds another keyword \emph{Large Language Models} to better reflect her research interest\minor{s}. She also notices a keyword \emph{multitouch technology}, which reveals an aspect she has \minor{not} thought of before. \sys{} saves the updated description and uses this description throughout the rest of the process.

After setting up the folder, the researcher receives a paper alert email when there is a new set of recommended papers. By clicking on the link in the email, she is directed to an interactive paper alert interface (Fig. \ref{fig:interface}). She bookmarks this link for future reference. She quickly goes through the recommended papers and immediately saves papers with obvious relevance (e.g., ``The Semantic Reader Project: Augmenting Scholarly Documents through AI-Powered Interactive Reading Interface''\cite{Lo2023TheSR}) based on the provided title and metadata (authors \minor{and} TL;DR). For \minor{a} paper with non-obvious connections, she explores the contextualized descriptions of the paper. She first opens the \textit{Problem, method, and findings} tab to see a summary of the paper by the folder context broken down into problem, method, and findings. For example, for the recommended paper ``Synergi: A Mixed-Initiative System for Scholarly Synthesis and Sensemaking''~\cite{synergi}, she can see that \sys-provided method of the paper is relevant to her interest in LLM ("Method: Developed Synergi, a mixed-initiative workflow tying user input, citation graphs, and LLMs") and the paper's problem is "Problem: synthesizing research threads from multiple papers is challenging". Seeing such connection\minor{s} with \emph{Reading-supporting interfaces}, she saves the paper to her library folder. Feeling curious about the Synergi paper, she decides to explore the paper-paper relationship descriptions under the \textit{Relate to Paper} tab. She chooses to compare the recommended paper with a paper ``CiteRead: Integrating Localized Citation Contexts into Scientific Paper Reading''~\cite{rachatasumrit2022citeread} that she is already familiar with. She learns that the system in the Synergi paper requires more active engagement from users compared to the system in the CiteRead paper. Now, she notes interactivity as one dimension \minor{of} the reading interfaces design space. She also sees that there is a description for the Synergi paper and a paper she has collected in passing. By reading the description, she now recalls that ``Scim: Intelligent Skimming Support for Scientific Papers''~\cite{Fok2022ScimIS} supports paper reading by automatically highlighting important content in a paper and, unlike Synergi, there is no personalization component in the system. Compared to other traditional paper alert systems that only show a list of titles and, at most, abstracts, \sys{} helps the researcher figure out how each of the recommended papers connects to her topic of interest, as well as their nuanced relationships to papers she had already collected.


\subsection{Methods for Generating Contextualized Descriptions}
\begin{figure*}[!h]
    \centering
    \includegraphics[width=\textwidth]{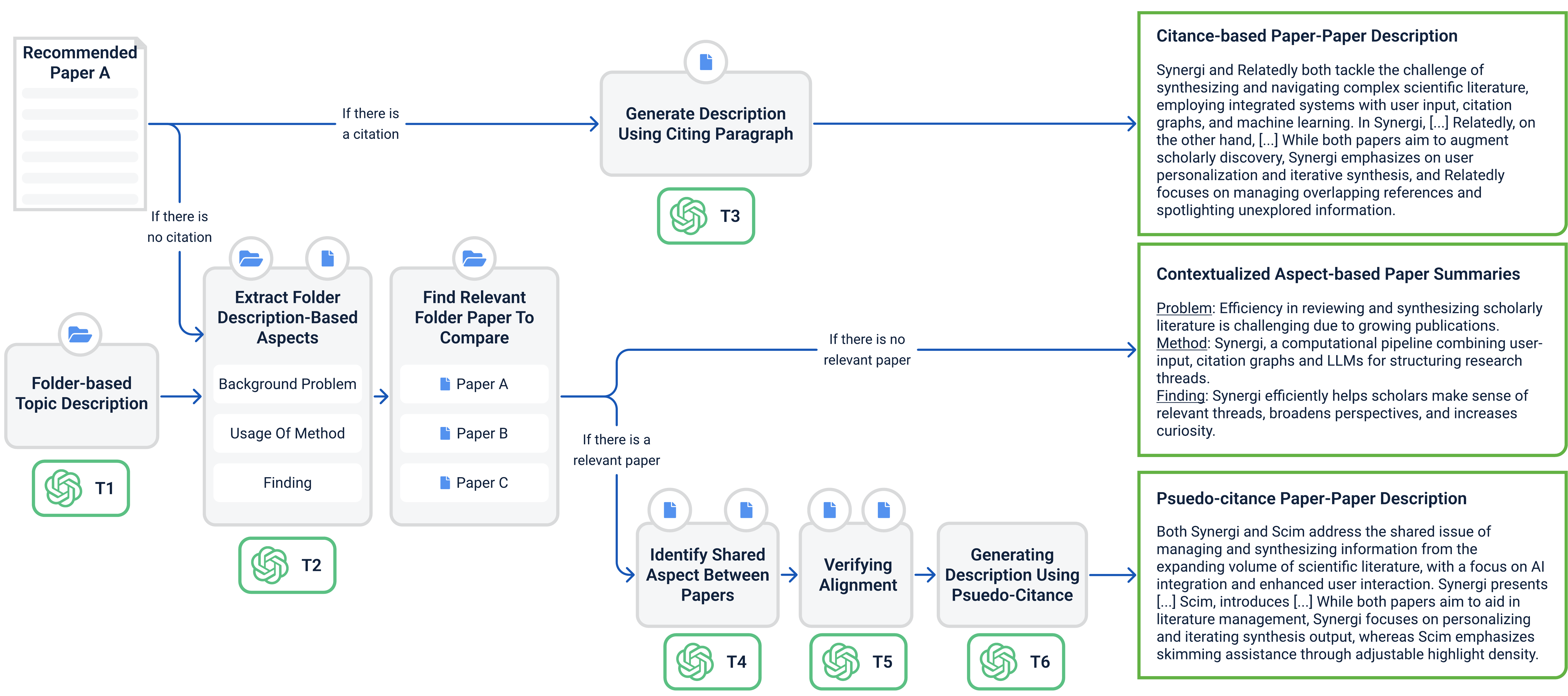}
    \caption{Overview of \sys{}'s pipeline to generate contextualized description\minor{s}. If there is a citation from the recommended paper to a collected paper, \minor{\sys{}} generate\minor{s} \textit{paper-paper descriptions based on citances} \minor{using citing paragraph}. If there are no citances, \sys{} synthesizes pseudo-citing sentences \minor{that shows the relationship between recommended paper and relevant collected paper to make} \textit{paper-paper descriptions via generated pseudo-citances}. If there are no \minor{relevant} collected papers, \sys{} generates \textit{contextualized aspect-based paper summaries} with the folder's overall context. \pipeline{Prompt T1-T6 are in the Appendix~\ref{appendix:prompts}.}} 

    \label{fig:method}
\end{figure*}

To generate contextualized descriptions like those presented in the scenario, we developed an LLM-based pipeline (Fig.~\ref{fig:method}) that processes the user's folder information and the recommended paper's content to generate descriptions for each recommended paper.
\rationale{LLMs can make meaningful improvements in comprehension and summarization, particularly for long, complex documents that demand a high degree of accuracy~\cite{claude}. This capability enables \sys{} to identify relevant aspects with the given context in papers and synthesize descriptions from multiple aspects.}
\sys{} generates three types of descriptions: (1) \textit{contextualized aspect-based paper summaries}, (2) \textit{paper-paper descriptions based on citances}, and (3) \textit{paper-paper descriptions via generated pseudo-citances}.
We describe how \sys\ generates a compact summary of collected papers in a topical folder (\S\ref{s:gen-topic}) and uses the folder summary to generate contextualized aspect-based summaries (\S\ref{s:gen-aspects}), contextualized paper-paper relationships when recommended papers cite one or more collected papers (\S\ref{s:gen-paper-paper-citance}),  and when they do not (\S\ref{s:gen-paper-paper-pseudo-citance}). \pipeline{Full prompts are in Appendix ~\ref{appendix:prompts}.}



\vspace{-0.2cm}
\subsubsection{Suggested Topic Description}
\label{s:gen-topic}
To ensure the three types of descriptions (detailed in the \minor{following} subsections) are relevant to the topic \minor{of the folder}, \sys{} allows users to provide a compact description for their folders in addition to \minor{the} folder name. This topic description is used in all subsequent LLM prompts to convey the user's interests.
To lower the effort of writing folder descriptions, \sys{} uses an LLM to generate a default description based on collected papers already saved in the folder. \rationale{We adapted our prompt design from the prior \textit{LAMP} approach\cite{salemi2023lamp}, which creates a user profile prompt with \minor{the information from the} user's own paper. Our prompt takes a task instruction and a list of titles of papers included in the folder as inputs} \pipeline{(T1 in Appendix ~\ref{appendix:generating_fd}).}
Then, our prompt instructs an LLM to generate a description including the folder title, shared goals of collected papers, and a list of topical keywords (Fig.~\ref{fig:interface}A; folder title and description). This default folder description was then shown to the user. The user could further edit them to reflect their interests beyond the papers that they had collected. 


\vspace{-0.2cm}
\subsubsection{Contextualized Aspect-based Paper Summaries}
\label{s:gen-aspects}
Among the various aspects of the recommended paper, we extract the aspects (\ie rhetorical structure elements indicating problem, method, findings \cite{Mysore2021CSFCubeA}) that \minor{the} user who has curated this library folder might \minor{find} relevant.
The problems, methods, and \minor{findings} are typically the main pillars of most papers \cite{chan2018solvent}. At the same time, these aspects can describe research in a specific and comprehensive way.
\pipeline{
To extract a set of problems, methods, and findings in the context of the library folder,
our method takes a title and an abstract of the recommended paper and the folder description that represents the user's research interest as inputs (T2 in Appendix ~\ref{appendix:generating_aspect}).} \pipeline{We guide an LLM to identify as many relevant problems from the recommended paper as possible. Then, we describe the specific methods applied by the paper for each problem and elaborate on the specific findings identified by applying each method.}
Our method uses these aspects not only as the summary of a single recommended paper, we further build on the extracted aspects to align two papers to create pseudo-citances as described in \S\ref{s:gen-paper-paper-pseudo-citance}.

\vspace{-0.1cm}
\subsubsection{Paper-paper Descriptions Based on Citances}
\label{s:gen-paper-paper-citance}
For this description type, we exploit citations between recommended papers and collected papers. Citation sentence\minor{s} (\ie citance\minor{s}) are widely used as prox\minor{ies} of the relationship between the citing and the cited paper\minor{s} \cite{Luu2020ExplainingRB}.
Among the intents of citances (\ie background, method, or result\minor{s}), citances with the background intent \minor{give} more context about a problem, concept, approach, topic, or importance of the problem in the field \cite{Cohan2019StructuralSF}.
\minor{Citances with the background intent} often contain information about how a citing paper present\minor{s} a new approach and how it compare\minor{s} to a cited paper. Since our formative study revealed that participants regarded ``build on'' relationship to be important, \minor{we prioritize background citances as classified by \cite{Cohan2019StructuralSF}} when selecting a citance to generate a description.

\rationale{\minor{A} citance \minor{by} itself is hard to understand for users without any context.} \minor{O}nce a citance was selected in \minor{a} recommended paper, we extracted the paragraph which \minor{the} citance was in (\ie citing paragraph) to obtain additional context around it.
\rationale{Rather than showing the citing paragraph that only has partial content of \minor{the} recommended paper,} we employ \minor{an} LLM to generate a compact but detailed description that describes both the recommended paper and its relationship to the cited collected paper \minor{as mentioned in the citing paragraph.}
Additionally, to support our DG3 of helping users learn new aspects of collected papers, we added a short summary of the cited collected paper.
\pipeline{To obtain this description, we designed inputs of the prompt to include the titles and abstracts of both the recommended and the cited paper, as well as the citing paragraph (T3 in Appendix ~\ref{appendix:desc_citance}).}




\vspace{-0.1cm}
\subsubsection{Paper-paper Descriptions via Generated Pseudo-citances}
\label{s:gen-paper-paper-pseudo-citance}
The method described in the above section (\S\ref{s:gen-paper-paper-citance}) does not apply to all recommended papers. Authors typically do not comprehensively cite all relevant papers. Some relevant papers are omitted from the citances even when they \minor{are} relevant to some of the collected papers. 
We build on prior work that used the problem-method-findings schema to show similarities and differences between papers in a search engine setting. Our method generates structured summaries using a similar approach to describe paper recommendations anchored to previously collected papers.
In our design, this type of description feature\minor{s} two core relations primitives between papers: 1) \textit{comparisons}, which provide concise descriptions of \minor{the} most salient similarities along either the problem \minor{or} the method aspect and 2) \textit{contrasts}, which surface differences along a different aspect (\eg two papers that tackled the same research problem but used different methods). This structure has been shown to facilitate scholarly sensemaking and inspirations (\S\ref{subsection:related_work_depth}.)

\rationale{Rather than concatenating each paper's aspects that are similar or different with extractive techniques, our method aims to provide well-aligned comparisons and contrasts between two papers. This method is motivated by the formative study where participants mentioned that simple concatenation was not helpful in making sense of the relationships between two papers.}
\pipeline{To create this description, our method follows the following process: (1) find relevant papers, (2) identify shared aspects for each pair of the recommended paper and a collected paper to find similarities and differences between them, (3) verify whether the shared aspect is aligned with both papers and (4) generate \minor {a} structured summary.

Here, we explain our approach in more detail, using the case where the recommended paper and the collected paper share similar ``problems''.}
Our approach \minor{first} selects \minor{the} top-5 most similar collected papers to a recommended paper based \minor{on} the abstract similarity using \pipeline{Flag embedding\minor{s}~\cite{bge_embedding}, a state-of-the-art text embedding model}. 
Then, with the aim of finding similarities and differences between papers, we \minor{use} \minor{the} method described in \S\ref{s:gen-aspects} to extract multiple problem-method aspect pairs from each paper's abstract. 
\pipeline{
By providing \minor{the} titles and \minor{the aspects} of all five relevant collected papers and the recommended paper as inputs, we instruct an LLM to (1) identify \minor{the} top-5 papers that have problems that are the most similar \minor{to the} problems in the \minor{recommended} paper, (2) list all of the identified pairs (one from the given paper and the other from a collected paper), and (3) describe one shared problem that could encompass the two identified problems.}
To \minor{confirm} whether the shared problem encompasses problems of both the recommended paper and the chosen collected paper, our approach prompt\minor{s} \minor{an} LLM to verify whether \minor{the} shared problem is addressed in each paper. \minor{This was done by providing each paper's title, abstract, and the generated shared problem.}
Finally, by inputting \minor{respective contrasting methods employed} in \minor{the} recommended and \minor{the} folder paper, along with the generated shared problem and \minor{the} abstracts of both papers, the LLM generates the structured summary. This summary includes \minor{comparing} and contrasting sentences with short summaries of two papers. While this process explains how to generate a description for two papers with similar problems, a similar process applies to a description of two papers with similar methods.

\vspace{0.2cm}

\subsection{Paper Alert Interface}

For each folder, \sys{} takes the folder title, the folder description, and a list of papers in the folder as inputs. We use the \minor{publicly} available Semantic Scholar Paper Recommendation API\footnote{\url{https://api.semanticscholar.org/api-docs/recommendations}} to retrieve a list of recommended papers for a given set of folded papers. 
For each folder, a dedicated web page allows users to view recommendations displayed on detailed cards, featuring information about the paper (title, authors, venue, and year) and a machine-generated TL;DR summary~\cite{Cachola2020TLDRES}.
The title of \minor{a paper links} to the corresponding paper details page on Semantic Scholar where users can access the pdf file of the paper (if available) together with other information such as the paper's citations and references. Users can explore different descriptions displayed in three tabs:
\begin{itemize}
    \item \textit{Related to Paper} (Fig. \ref{fig:interface}B) shows paper-paper descriptions that \minor{are} focused on showing relationships between the recommended paper and a specific \minor{paper} previously saved in the folder. The descriptions shown in this tab are \textit{paper-paper descriptions based on citances} and \textit{paper-paper descriptions via generated pseudo-citances}. The two papers mentioned in the text are highlighted in different colors to indicate which paper the description refers to. Users can use the dropdown control to select descriptions for the relationships between the recommended paper and a specific paper.
    \item \textit{Problem, method, and findings} (Fig. \ref{fig:interface}C) shows the \textit{contextualized aspect-based paper summary} for the recommended paper broken down into three aspects.
    \item \textit{Abstract} (Fig. \ref{fig:interface}D) shows the original full abstract of the paper.
\end{itemize}

Through the paper alert interface, users can save a recommended paper (Fig. \ref{fig:interface}E) to their folder and write a note (Fig. \ref{fig:interface}F) about the recommended paper for future references.

\subsection{Implementation Details}
\sys{} interface is a standard web application. The back-end was implemented in \minor{Python} using Flask for an HTTP server and PostgresSQL for \minor{a} database. The front-end was written in TypeScript using React framework. \sys{} retrieves the paper's metadata (title, authors, abstract, etc.), TL;DR and citances from the public Semantic Scholar API \footnote{\url{https://api.semanticscholar.org/api-docs/graph\#tag/Paper-Data/operation/get_graph_get_paper}}. We obtain the extracted plain text for papers using S2ORC, an open-source PDF-to-text extraction pipeline and a corpus of processed 81.1M academic papers across multiple disciplines \cite{Lo2020S2ORCTS}. We use GPT4-0613 through OpenAI API\footnote{\url{https://api.openai.com/v1/chat/completions}} for all text generation with \minor{an} LLM.


\section{User Study} \label{05_user_study}
We conducted a within-subjects laboratory study to investigate whether \sys{} can help readers efficiently understand the relevance of new papers, triage them, and deeply understand their relevance to users' own context.
\subsection{Research Questions}\label{subsection:RQs}
To this end, we formulated our research questions as follows:
\begin{itemize}
\item[\textbf{[RQ1]}] How do relevance descriptions augmented by user-collected papers’ information help a user understand how recommended papers are relevant to the user’s research context?
\item[\textbf{[RQ2]}] How do relevance descriptions augmented by user-collected papers’ information help a user triage the paper?
\item[\textbf{[RQ3]}] How do relevance descriptions augmented by relationships help a user understand the relationship between user-collected and recommended papers?
\item[\textbf{[RQ4]}] How do relevance descriptions augmented with a set of user-collected papers help a user understand new aspects of the collected papers?
\end{itemize}

\subsection{Study Design}
\vspace{0.2cm}
\subsubsection{Participants}
We recruited 15 researchers who are Ph.D./MS students in the CS domain for the study through snowball sampling and the authors' social media (Twitter). Two of the participants were Master's students, seven were junior (first- and second-year) doctoral students, six were senior (\minor{third-}, fourth-, and fifth-year) doctoral students. Our study focused on Ph.D./MS students as they may receive the greatest benefit from explaining why those papers are connected to their personal research context. Ph.D./MS students need to keep up with the large amount of literature the most. All participants reported that they read paper alerts, from monthly to daily. 

\begin{figure}
    \centering
    \includegraphics[width=\linewidth]{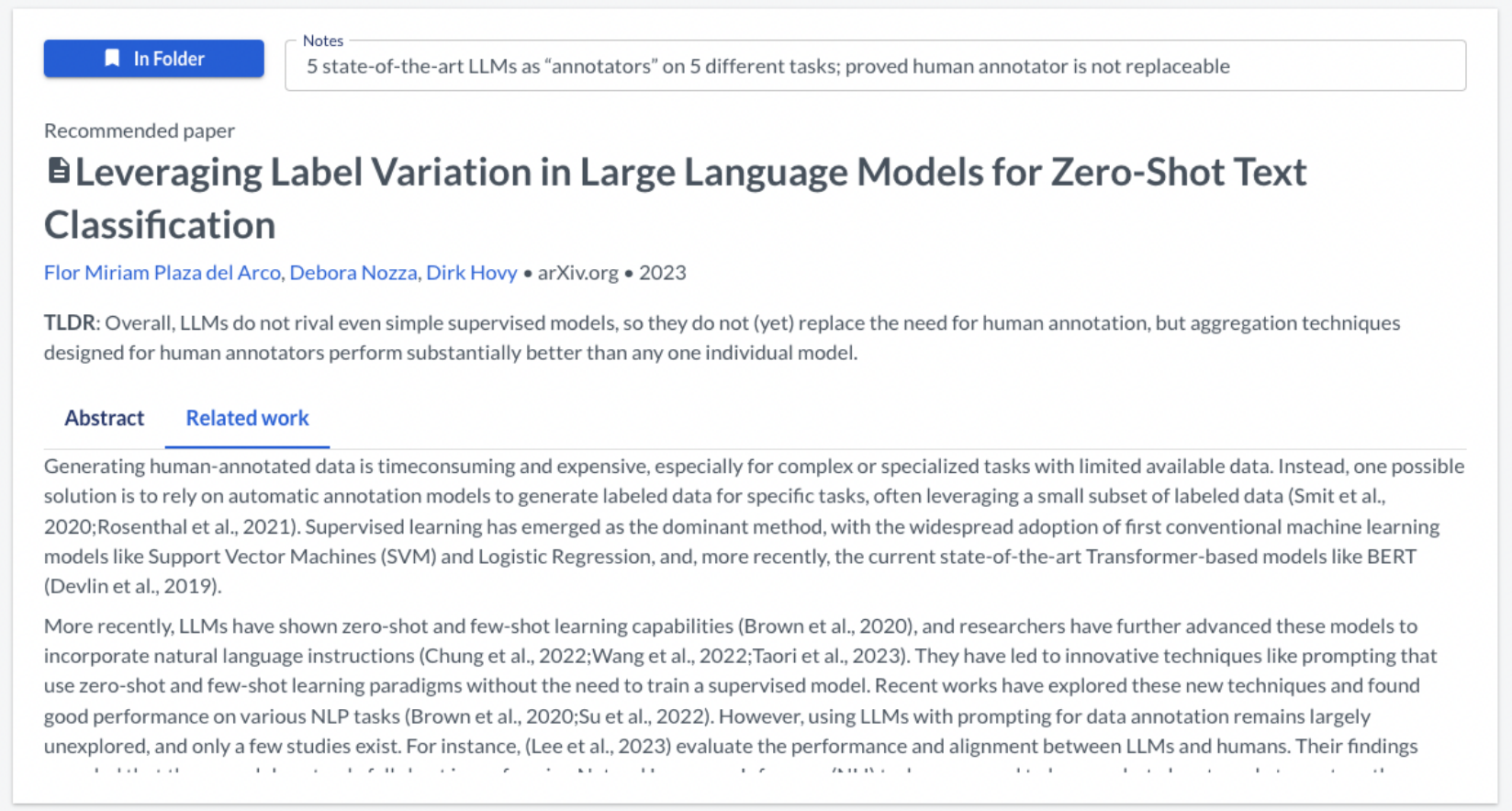}
    \caption{In the baseline condition, participants also see the recommended papers in a similar paper card to \sys{} paper alert interface. Instead of the three tabs with different types of explanation, there are two tabs: \textit{Abstract} that shows the abstract of the recommended paper and \textit{Related work} that shows the text of the related work section of the recommended paper. The rest of the functionalities remain the same across conditions.}
    \label{fig:baseline}
\end{figure}
\subsubsection{Conditions}
For the study, participants read paper alerts and listed what they learned from them using either \sys{} or a baseline system. Like in \sys{}, each recommended paper in the system for the baseline condition (Fig \ref{fig:baseline}) contains its TL;DR, abstract, and the link to the paper's details (the paper details page on Semantic Scholar). Additionally, we provide excerpts of the related work section of the recommended papers in the baseline condition. The related work section usually shows relationships between the paper and prior work. Researchers are likely to check this section to look for relationships between the recommended paper and other papers. While both \sys{} and the baseline system give the link to the full paper to inspect the paper further, having the related work section on the baseline condition helps balance the amount of readily available information between the two conditions.
In the \sys{} condition, \textit{\minor{contextualized} aspect-based paper summaries} and \textit{paper-paper descriptions} are additionally provided. 
The ordering of the conditions was counterbalanced to mitigate potential ordering effects.

\subsubsection{Paper Alert Setup}
We asked participants to create their own library folders (paper collections) on Semantic Scholar and save papers into them before the study. 
To balance the quality and familiarity of folder content between the conditions, we asked participants to follow the instructions:
\begin{itemize}
\item[1.] (Scenario and Number of Topic Folders) Pick three research topics for curation, and fix the use scenario (\eg surveying the literature to write a related work section in your paper).
\item[2.] (Topic Familiarity) Topics should be similarly familiar to you.
\item[3.] (Number and Familiarity of Papers) Collect five papers you fully read, three papers that you have read somewhat, and two papers that you have seen but not read.
\end{itemize}
We chose two folders from the three participants brought with them based on the following criteria, to reduce the significance of potential quality and familiarity differences between the papers actually used in the experiment: 
We generated recommendations for each folder and ranked them based on the content similarity using the average of cosine similarity between library papers' and recommended papers' SPECTER~\cite{cohan-etal-2020-specter} embeddings; pick the top two.
Regarding folder descriptions, participants saw an example and edited them before the study.

\subsubsection{Procedure}
The study was conducted remotely using video conferencing software.
After a brief introduction to the overall study, participants were given a short tutorial of the system (either treatment or control, counterbalanced).
They interacted with the interface for 2 minutes to familiarize themselves with the interaction features using fake data. 
After the tutorial, participants were given 12 minutes to read the descriptions on 8 recommended papers.
After they read the descriptions, they were provided with survey questions.
Once the survey was completed, participants were given 10 minutes to write down a list of what they learned from reviewing the descriptions and wanted to retain for future use.
Participants were asked to use their own language to describe the things and were informed the quality of writing was irrelevant. The steps above were repeated for the second condition. Finally, the experimenter conducted a semi-structured interview about participants' overall experience. The study lasted for $\sim$90 minutes, and participants were compensated \$80 USD for their time. The study was approved by our internal review board.

\subsection{Measures}
\subsubsection{Overview} In relation to the research questions described in \S\ref{subsection:RQs}, we measured participants' subjective responses to corresponding survey questions.
To triangulate the survey responses, two of the authors manually coded what participants wrote down as their learning from reading the descriptions in terms of whether it contained elaboration on the relationship between a recommended paper and the saved papers participants began with or their curiosity for further investigation.
We also employed a survey asking participants' general experience with each system.
Finally, we conducted a sanity check on the degree of factual hallucination among the descriptions generated in \sys, to quantify its current limitations and point to future work directions.

\subsubsection{Targeted Survey Responses} \label{subsubsection:targeted_survey} We collected participants' responses (7-point Likert scale) to the following five questions from the survey: ``\emph{The system helped me understand how the recommended papers are relevant to my research interest.}'' (RQ1), ``\emph{The system helped me decide which papers are worth saving.}'' (RQ2), ``\emph{I was confident in deciding which papers were relevant and worth saving.}'' (RQ2), ``\emph{The system helped me understand how the recommended papers relate to the papers I have already saved.}'' (RQ3), and ``\emph{The system helped me learn something new about the papers I have already saved.}'' (RQ4).

\subsubsection{General Experience Survey Responses} \label{subsubsection:nasa_tlx} To measure perceived workload, the survey also included five questions (excluding the physical demand question) from the NASA-TLX questionnaire~\cite{nasa_tlx}, where a more compact 7-point scale, mapped to the original 21-point scale, was instrumented~\cite{nasa_tlx_7point_use_case}.

\subsubsection{Annotating Participants' Learning} \label{subsubsection:annotation_measures} To deeply analyze participants' learning from reviewing the descriptions, we constructed an annotation regime in five dimensions to capture the quantity of facts and relationships described in participants' notes, and the level of elaboration around relationship- or curiosity-related details.
Specifically, the five dimensions were: 1) the number of facts about papers; 2) the number of groups of papers synthesized by participants; 3) the number of relationships described between a paper and a group of papers or between groups of papers; 4) the level of factual details on the relationships; 5) mentions of curiosity and/or authentic motivation for further investigation.
The first three dimensions were coded using an interval scale, ranging from 0; the last two dimensions were coded using a binary scale, where 0 represented `lacking elaboration'.
We sampled a set of randomly sampled 15 notes participants wrote from the full set of 169 notes for two annotators (two of the authors who were not experimenters in the study).
The annotators coded and discussed these samples over two rounds to reach a consensus. 
The following is the rubrics developed for annotation:
\begin{itemize}
\item[\emph{Step 1}] Identify a paper or a group of papers described in the comment.
\item[\emph{Step 2}] Identify whether there is a relationship described between a paper and a group of papers, among a group of papers, or between groups of papers.
\item[\emph{Step 3}] Once at least \emph{one} relationship is found, find whether there are factual details or concepts elaborated about the relationship in the comment.
\item[\emph{Step 4}] Similarly with \emph{Step 3}, identify if participants expressed surprise, curiosity, or motivation for future investigation in the text (\eg ``I'm not sure how that difference manifests yet.'', ``I'm surprised by the results, I'll look into the papers later.'').
\item[\emph{Step 5}] Count the number of distinct facts, using the clause or sentence boundaries.
\end{itemize}
The annotators annotated 20 shared samples (randomly drawn) blind-to-condition to check for inter-rater reliability.
This resulted in Krippendorff's $\alpha$'s ranging between 0.46 (the binary measure of factual detailedness in descried relationships) and 0.78 (the number of groups of papers synthesized by participants).


\subsubsection{Accuracy of LLM-generated Descriptions} Additionally, as LLM can generate hallucinations (\ie factually incorrect information), we collected annotations of factual correctness for 60 descriptions for each of \pmf{} and \rel{} used in the user study. Since evaluating the factual correctness of our description requires expertise in these domains, we recruited two experts per domain for the HCI and NLP domains for annotation. Each expert annotated the data for recommended papers in their domain of expertise. For each description, experts rated its factual correctness on a three-point scale based on counting the number of factually incorrect mentions in descriptions (3: None, 2: 1-2, 1: 3 or more) and wrote how they are incorrect. Two experts in each domain independently annotated 40 descriptions per expert (20 per each type), where 20 descriptions get two annotations (60 samples are evaluated in total per domain). Krippendorff's $\alpha$ is 0.55 for HCI paper descriptions and 0.41 for NLP ones. 




\subsection{Analysis}
We analyzed the survey data (\S\ref{subsubsection:targeted_survey} and \S\ref{subsubsection:nasa_tlx}) by first conducting the Shapiro-Wilk test to determine if the data required a parametric (P) or non-parametric (NP) testing procedure and proceeded with a paired t-test (when parametric) and a Wilcoxon signed-rank test (when non-parametric) accordingly. In order to account for the between-subject variability in participants' notes data (\S\ref{subsubsection:annotation_measures}), we employed a Mixed Linear Model Regression. The dependent variable in the model was each of the five dimensions of annotations, random effects were the participant IDs, and the fixed effects in the model were the experimental conditions.
We used Restricted Maximum Likelihood (REML) for parameter estimation to ensure robust estimates of the variance and covariance parameters in cases when the homoscedasticity and normality of data assumptions are violated.
We report on the results of the regression analyses as regression coefficients ($\beta$'s) and $p$-values, with significance indicated at the $\alpha=.05$ level. We also report the fixed effects sizes (semi-partial R$^2$, which represent the \% outcome variance, controlling for the predictors and random effects in the model) when applicable.
\section{Findings}
Our results showed that \sys{} helped participants understand the relevance of recommended papers and make more effective relevance judgments. Additionally, \sys{} promoted the discovery of more connections between papers that were relevant to the topic of the folder, and led to writing more detailed notes that contained rich connections between papers.

\subsection{General Behavioral Differences}
During the study, we asked participants to think-aloud and observed how they interacted with the two systems.
In both conditions, participants typically did a quick pass over the recommended paper list to filter out few recommended papers that seemed obviously irrelevant. When interacting with the baseline system, participants relied on the titles and the TL;DR summaries for this quick triaging process, while when interacting with \sys{} they additionally considered the \pmf{}. 
P2 mentioned that even though the two summaries were similar in length, TL;DR summaries typically focused on the research problems that were not always relevant to their folder's topic. In contrast, the \pmf{} provided by \sys{} often surfaced parts of the abstracts that were relevant.
At the same time, participants in both conditions said that after this first pass they still needed to examine most of the papers more carefully to understand how they are relevant and to confidently judge which ones to save.

When interacting with the baseline system, participants continued to read the full abstract or open the papers to see its figures but said this process was effortful. For example, P9 said that \textit{``I tried to understand [the recommended] papers but reading all of the abstracts is overwhelming.''} Further, even when they tried to carefully examine the recommended papers in the baseline condition, they often failed to identify the connections.
For example, P1 mentioned that \textit{``I cannot understand why this paper is recommended to me. It seems they are talking about just their own topic''} and P15 said that \textit{``Abstract alone cannot answer `how does this paper relevant to my research context?'}.

In contrast, when interacting with \sys{}, participants relied on \rel{} after this first pass. In this case, many participants mentioned that they can better understand \textit{``how this paper is different from the paper that they've already known''} and \textit{``what new contributions are there [in the recommended papers]''} that were relevant to the collected paper. Interestingly, after reading the \rel{}, participants often still continued to examine the abstracts. Participants mentioned the goal was to both verify the LLM-generated \rel{} with the original sources and to ``gain deeper context'' around the \rel{} once they became interested in a recommended paper (more details around this behavior in \S6.4).

\subsection{Exploring Papers Broadly by Understanding Relevance}
\begin{figure*}[!t]
    \centering
    \includegraphics[width=1.00\textwidth]{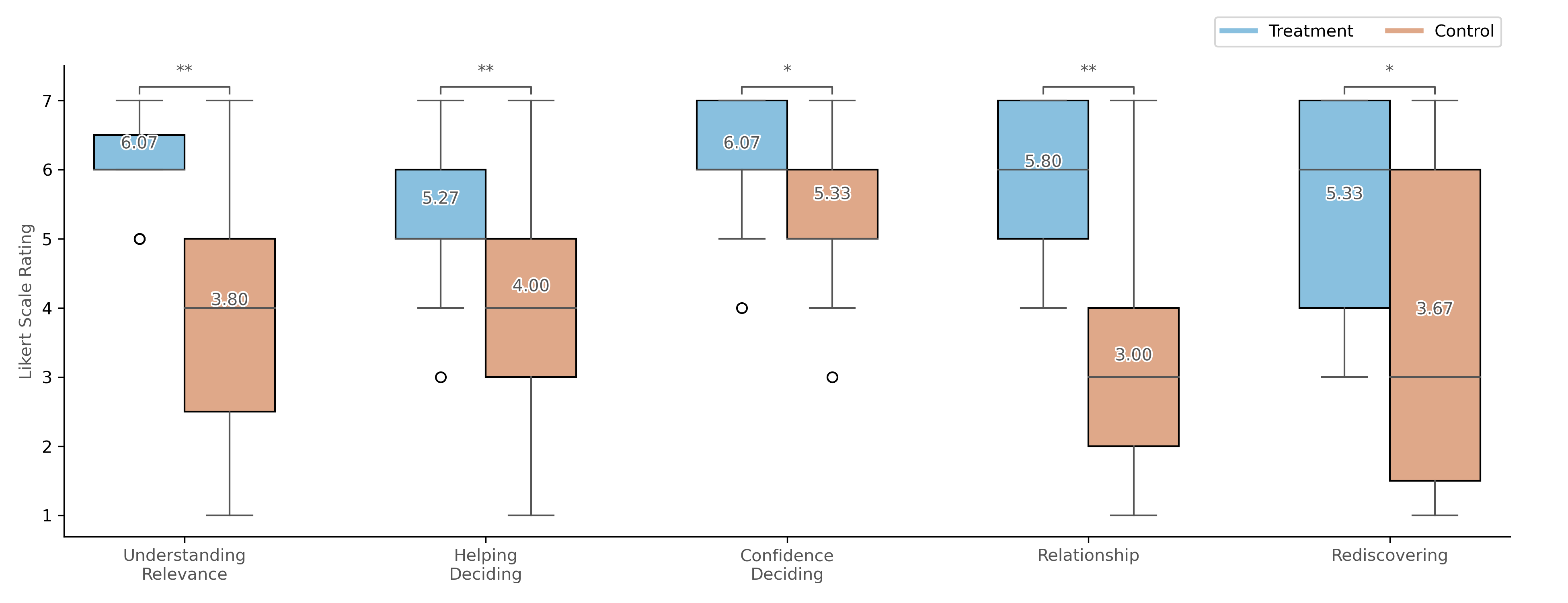}
    \caption{Results of the post-survey showed that participants found various benefits when using \sys{} compared to a strong baseline. **, *, and ns indicate significance of $p < 0.01$, $p < 0.05$, and $p > 0.05$, respectively.}
    \label{fig:survey}
\end{figure*}

Based on the post-survey, participants felt that they could understand how the recommended papers were relevant to their own research interest significantly better in \sys{} ($M=6.07$, $SD=0.68$) than the baseline ($M=3.80$, $SD=1.64$, $p =0.0013 $, NP).
We also found evidence that \sys{} supported decision-making. Specifically, participants felt \sys{} helped them decide which papers were worth saving (\sys{}: $M=5.27$, $SD=1.06$, baseline: $M=4.00$, $SD=1.55$, $p = 0.0024$, P) and were significantly more confident in their decisions (\sys{}: $M=6.07$, $SD=0.85$, baseline: $M=5.33$, $SD=1.07$, $p = 0.0124$, NP).
Qualitative insights revealed that these perceptions were due to how \sys{} contextualized explanations for each user based on their topical folder.
According to our participants, both \pmf{} and \rel{} effectively highlighted which parts of the abstracts or papers they should focus on (P9, P13).
Specifically, while \pmf{} surfaced \textit{``explicitly relevant aspects dispersed in multiple sections [in the paper, relevant] to my folder context''}. Participants also appreciated how \rel{} connected recommended and collected papers with detailed and insightful relationships with \textit{``narrower research interest perspective''} when compared to uncontextualized TL;DR summaries. For example, P15 mentioned that \textit{``I don't have time to read all the abstracts but TL;DRs are too high-level. So with the [baseline], I do not have evidence to choose what to save. With the descriptions in \sys{}, I was more confident in my decision because I can get more understandable evidence from the explanations. For example, how the recommended papers [were relevant but] tackled different problems than the paper I knew already)''}.

\subsection{Deeper Insights on Both Recommended and Previously Collected Papers}
\begin{figure*}[!t]
    \centering
    \includegraphics[width=1.00\textwidth]{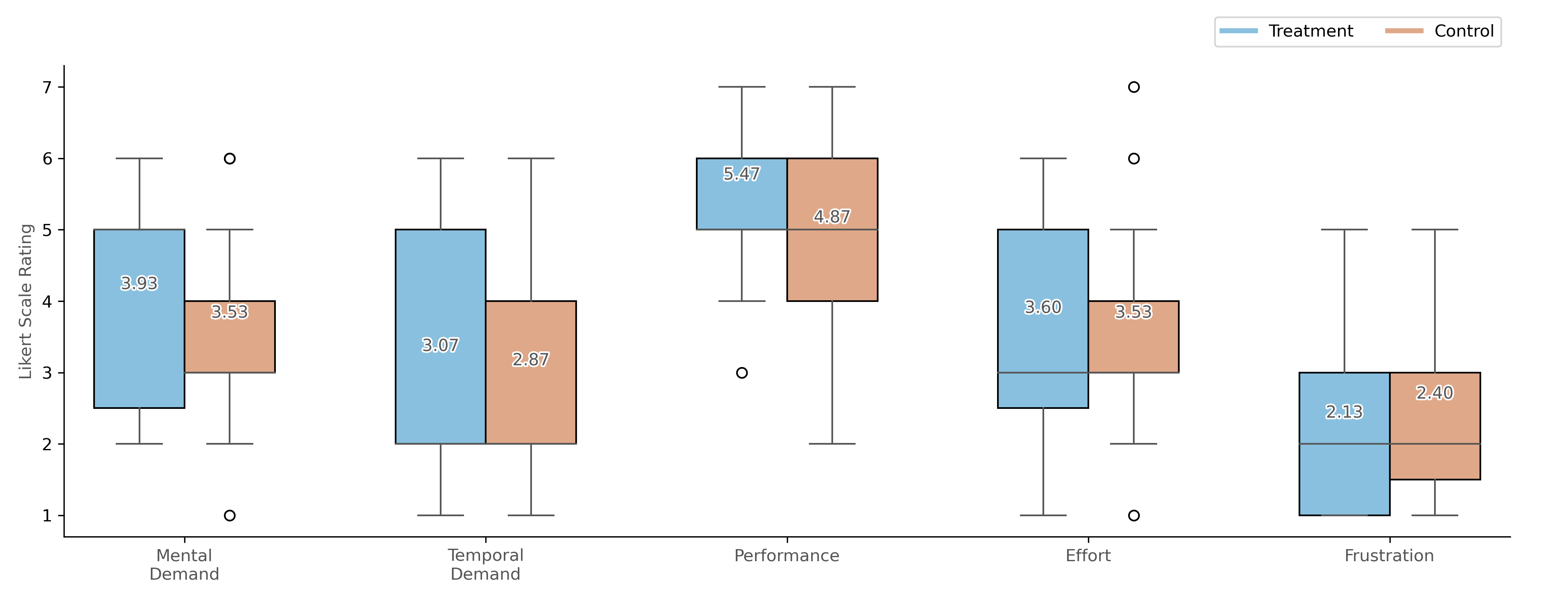}
    \caption{Based on a NASA-TLX survey, participants perceived similar workload when using \sys{} and the baseline conditions. **, *, and ns indicate significance of $p < 0.01$, $p < 0.05$, and $p > 0.05$, respectively.}
    \label{fig:nasatlx}
\end{figure*}
To gain a deeper understanding of the knowledge participants had gained from interacting with \sys{} and the baseline, we further analyzed the notes participants took during the study in the two conditions (while being blind to which conditions the notes came from). 
We found that when using \sys{}, participants on average captured significantly more notes that described connections between papers (\sys{}: $M=2.21$, $SD=1.81$, baseline: $M=1.07$, $SD=0.96$, $p=0.0459$, NP).
At the same time, when describing the relationships, participants included similar levels of details
after accounting for between- and within-participant variability ($\beta=.113$, $p=.13$). This suggests that the bottleneck of learning connections between papers is a recall problem, and that it is easier to capture these connections with \sys{} compared to looking at multiple paper abstracts in the baseline condition.
These results based on participant behavior also corroborate their perceived understanding of the information that was presented in the two conditions. Specifically, in the post-survey, participants reported that they could understand how the recommended papers connect to their collected papers significantly better with \sys{} (\sys{}: $M=5.80$, $SD=1.04$. baseline: $M=3.00$, $SD=1.67$, $p=0.0021$, NP).

One possible trade-off for the positive learning effect could require significantly higher cognitive demands on the users. However, based on our NASA-TLX survey, participants did not report a higher workload when comparing using \sys{} and the baseline.
On the other hand, qualitative data do show anecdotal evidence of higher mental and temporal demand for some participants when using \sys.
For example, P6 and P9 pointed to descriptions from \sys{} that peaked their interest in the recommended papers and urged them to explore them in more detail. P4 further commented that \textit{``I felt mentally more demanded using \sys{} because I actively compare the new papers with other [collected] papers. This might be a positive side effect of showing relationships.''}
These comments suggest that even for participants who felt using \sys{} was more cognitively demanding, they perceived it positively because it prompted them to become more actively engaged with the paper alert, motivating them to more deeply process the information in front of them.

Finally, in addition to understanding the recommended papers, our analysis of the post-survey data also revealed that participants felt \sys{} helped them \minor{\textit{``learn something new about the papers that they have already saved in the folder''} }significantly more than the baseline (\sys{}: $M=5.33$, $SD=1.49$. baseline: $M=3.67$, $SD=2.18$, $p=0.011$, NP). In the interviews, participants pointed to different benefits when seeing descriptions that covered relevant collected papers in the folders, including refreshing their memories about papers they had previously read and gaining new knowledge and perspectives about them based on how recommended papers described them in their related work sections. In addition, we also found some participants who had saved papers in their folder that they had not yet read, in this case, they describe how the descriptions from \sys{} helped them \emph{rediscover} previously collected papers with a \emph{renewed interest.}

\subsection{Accuracy of LLM-Generated Descriptions}\label{eval:quality}


One known issue with current LLMs is that they can be prone to hallucinations (\cf~\cite{thorp2023chatgpt,bang2023multitask,black2022tweet}), and we did find that some descriptions contained one or two mistakes based on manual evaluation. In general, \pmf{} are more extractive and contain fewer mistakes than \rel{} which relies on training knowledge to align the extracted aspects.
Specifically, for \pmf{}, 8\% of samples has one or two mistakes (HCI: 6\%, NLP: 10\%, based on 30 samples each), and \rel{}, 20\% (HCI: 16 \%, NLP: 23\%, based on 30 samples each). One fundamental question here is that - is this an acceptable level of accuracy in the context of paper alerts? Results from our user study around participants' strategies when interacting with \sys{} offered some insights.

Firstly, like most retrieval systems, results from document recommender systems can also contain errors (i.e., documents falsely classified as relevant) \cite{cohan-etal-2020-specter}. Because of this, we found that when interacting with paper alerts, participants were already in the mindset of verifying the machine-generated document recommendations, in both conditions. We also found participants go through the list of recommended papers in multiple passes, typically using the titles first to rule out clear irrelevant recommended papers, then reading the generated descriptions in the second pass, and finally verifying them with the abstract or the content of the papers. For example, representative quotes from P6 and P8 mentioned \minor{\textit{``validate the connections between them [papers] in the abstract''}} and P2 provided more details around their strategy: \minor{\textit{``[I read abstracts] not only to see deeper levels of information but also to verify the content of the descriptions.''}}

Secondly, many participants also explicitly mentioned that they see the \sys{} descriptions as ``\emph{supplementary material}'' that help them triage the recommended papers, suggesting that participants could appropriately adjust their levels of reliance on the LLM-generated descriptions. P2 said \minor{\textit{``I used the descriptions from [\sys{}] as supplementary material going from titles and TL;DR to abstract,''}} representative quote from P8 additionally pointed to benefits around awareness and discovery \minor{\textit{``\sys{}'s descriptions triggered me to become interested in two highly relevant [recommended] papers and have curiosity about them so that I can [decide to] read the abstracts and [then the] whole papers to get more information. These descriptions are supplementary bridges from the titles and TL;DRs to the abstracts rather than replacing them.''}}

Finally, results based on post-survey and NASA-TLX questionnaire suggested that \sys{} did not increase perceived workload even though participants were actively verifying LLM-generated content and that they were able to judge the relevance of the recommended papers more confidently and captured richer relationships between papers in their notes. We acknowledge that our participants were computer science researchers who currently might be more familiar with recent advancements in LLMs than researchers in other domains, although the increasing popularity of LLM-powered end-user tools also provides them increasing opportunities to interact with LLMs in other scenarios. Nevertheless, our results suggest that interfaces that make use of LLM-generated text should always provide both adequate indication when generated text is presented to its users and allow users to freely turn on or turn off generated content. \discussion{Moreover, it is essential to design effective mechanisms for users to verify content efficiently. Inspired by AngleKindling's system design~\cite{petridis2023anglekindling} that shows the connection between LLM-generated angles and source material to support journalists, one approach involves making specific sections of the paper, particularly relevant to the LLM-generated descriptions, accessible—such as through highlighting or indicating whether there is clear citations with a hyperlink to source papers. We also can indicate whether there is clear source material that the output is grounded on by marking citations at the end of each sentence.}

\section{Limitations and Future Work}
\subsection{\limitation{Limitations}}
\limitation{Our work has some limitations. We conducted our study with computer science graduate students who might not represent the broad spectrum of academic domains. While our approach is motivated by a general use case and designed to accommodate a broad range of academic domains, further evaluation with researchers from other academic domains, especially less technology-oriented ones, can help us understand how the generated descriptions can be applied broadly. Another limitation is our use of the Problem-Method-Findings schema. While this schema fits a large portion of papers, it might not cover some other papers such as survey papers or systematic review papers~\cite{higgins2008cochrane}. Further, researchers might want to apply their own schema (e.g., medical researchers would be interested in specific aspects of clinical trials). We discuss how to extend the schematic digests beyond Problem-Method-Findings in \S\ref{sec:extending_schema}. 
Finally, the measurements in our user study are more focused on subjective measures because 
it is inherently challenging to get robust and valid measures that score participants' understanding of the relevance between papers and measure their triaging performance due to the evolving notion of relevance personalized to each individual. 
Standardization of this notion would require careful examinations of prolonged interaction scenarios beyond the scope of the laboratory experiment conducted here.
While we obtained objective metrics in the experiment that can be evaluated independently of each researcher's context, as a future work, conducting a longitudinal deployment study could involve participants assessing their own written outputs over time, allowing for evaluations that consider evolving individual contexts.}

\subsection{Pairwise vs. Multiple-Paper Descriptions}
In this work, we focused on describing pairs of recommended and collected papers to help users contextualize unfamiliar papers with familiar papers. In our current design, each description only covers two papers: a recommended paper and a collected paper. This design decision was based on our formative study where we observed that participants with different familiarity with the topics all found benefits in seeing descriptions that covered two papers, but less experienced participants felt that descriptions covering three or more papers were too complex for paper alerts. 
Future work could explore ways to allow users to further customize their paper alerts by adjusting the complexity of the descriptions. One opportunity for future work in this direction is to look to research that focuses on automatic related work section generation~\cite{relatedly} or multi-document summarization~\cite{Giorgi2022TowardsMS} that aimed to generate descriptions for many documents. One interesting and relevant use case we observed in the user study was one participant who had saved a survey paper in their folder. In this case, the description \sys{} generated allowed participants to compare and contrast recommended papers with different research threads that were described in the survey paper's abstract and was perceived positively by the participants.

\subsection{Longer-Term Usage and Evolving Folder Descriptions}
In the user study, we observed that participants collected richer notes that captured information that connects multiple papers as opposed to about a single paper. Since \sys{} leverages a user’s folder name and description in its prompts to generate a description that can better reflect a user’s knowledge about the folder topic, an interesting future direction is to allow users to update their folder name and descriptions based on what they had learned in a paper alert. This information can be used to update the folder description to represent the user’s current knowledge and this can, in turn, improve subsequent paper alert generations. In this sense, the user’s folder can serve as an evolving external representation of the user’s understanding on a specific research topic. However, the longer-term effects of accumulating additional notes need further investigation.

\subsection{Extending Schematic Digests beyond \emph{Problem-Method-Findings}}\label{sec:extending_schema}
Participants in our studies have also commented on how they would like to further customize the information presented in paper alerts, pointing to avenues for future work.
First, some participants commented that they wanted to be able to surface information along certain other aspects of the schema, such as differences or similarities in evaluation regimes and their outcomes (\eg positive, negative); what types of study designs were run and how they were conducted (\eg controlled lab studies, field deployment studies, RCTs); approaches to developing AI models in a given problem domain; and the design of interaction features in proposed systems.
Such aspects of schemas were customized to different participants, suggesting that while the default information provided in \sys's problem-method-findings schema served as a useful entry into recommended papers, a deeper inspection following users' triage would benefit from further digesting papers along user-defined secondary schemas.

However, the first-and-secondary division of schemas that adapts to users' interaction with paper digests over time also suggests that there is an important gradient of specificity that may have to be designed to be adaptively adjustable based on user interaction, for example by utilizing a form of passive sensing over users' intent based on their interaction.
Clear examples of this are when participants stated that they wanted to ``see more'' in the aspect-based paper summaries, demonstrating an intent for lower-level details; another intent that participants may express would be wanting to ``adjust'' an explanation provided for a recommended paper, for example, as pointed out by P4, when the problem-method-findings schema in our approach did not provide useful information for survey papers because the problem and method descriptions were presented at too high of a level, abstracting away useful details of any individual papers or groups of papers synthesized by the authors of the survey paper.

\subsection{Contextualized Paper Descriptions beyond Paper Alerts}
Finally, while we focused on the scenario of helping users make sense of paper recommendation alerts, the proposed pipeline can potentially be generalized to other scenarios where users need to make sense of unfamiliar papers. For example, using one's publications as ``collected papers'' and generating descriptions for papers from another author to explore common research interests and facilitate collaborations. Another opportunity is to enrich a user's experience when \emph{reading} a related work sections in a paper (\cf CiteSee~\cite{chang2023citesee}, Threddy~\cite{Kang2022ThreddyAI}, CiteRead~\cite{rachatasumrit2022citeread}) by generating alternative descriptions about the cited papers based on papers already familiar to the current user. Beyond comparing with existing papers, users can also compare their own draft of a paper with new papers that might be relevant to them to get insights or new perspectives for framing the contrast and comparisons between papers when organizing related work sections.

\section{Conclusion}
This work presents \sys{}, an enriched paper alert system that provides contextualized text descriptions of recommended papers based on user-collected papers.
\sys{} leverages an LLM-based computational method to infer users' research interests from their collected papers and extracts contextualized aspects in recommended papers that are relevant to the inferred user's interest. Our method establishes relationships between recommended and collected papers by comparing and contrasting these contextualized aspects. Through a within-subjects user study ($N=15$), we found that \sys{} helped researchers more confidently make sense of paper recommendations and discover more useful relationships between recommended and collected papers in the folder when comparing with a baseline interface that enriched existing paper alert systems with abstract summaries and extracted related work sections.

\begin{acks}
The authors would like to thank Daniel S. Weld, Doug Downey,
and the researchers in the Semantic Scholar team for their insightful feedback and all the support for this work. 
We also thank Benjamin Newman and Tae Soo Kim for their thoughtful discussions.
We also thank the anonymous reviewers for their constructive feedback. 
Finally, we would like to thank the various researchers who participated in our pilot test and user study.
\end{acks}

\bibliographystyle{ACM-Reference-Format}
\bibliography{references}

\appendix
\appendix

\section{Prompts}
\label{appendix:prompts}
All prompts used in \sys{} are listed below. The {\textcolor{blue}{blue text}} represents the input content.

\subsection{[T1] Generating Folder Description}\label{appendix:generating_fd}

\textbf{System Prompt} 
\texttt{You are an intelligent and precise assistant that can understand the contents of research papers. You are knowledgeable in different fields and domains of science, in particular computer science. }

\noindent\textbf{User Prompt}
\texttt{This is my scholarly library, titled \textcolor{blue}{folder title}. The following papers are included. Write down two-line descriptions about this library that deal with high-level characteristics of these works commonly shared. Present the result as "Title: <given title>; Description: <two-line descriptions starting with "It encompasses">.\\
\\
{[Library papers]}\\
\textcolor{blue}{A set of titles of library papers}
}

\subsection{[T2] Generating Contextualized Aspect-based Paper Summaries}\label{appendix:generating_aspect}

\textbf{System Prompt} 
\texttt{You are an intelligent and precise assistant that can understand the contents of research papers. You are knowledgeable in different fields and domains of science, in particular computer science. You are able to interpret research papers based on the user's perspective.}

\noindent\textbf{User Prompt}
\texttt{
    We would like you to extract the dimensions of the paper based on my research interest. You will be given my research interest and a paper and will be asked to extract the problem, method, and findings that I might have interest in from the paper. You will be provided with the title and abstract of the paper and my research interest that describes the topics that I'm currently interested in.\\
    \\
    {[The Start of My Research Interest]}\\
    \textcolor{blue}{folder description}\\
    {[The End of My Research Interest]}\\
    \\
    {[The Start of Given Paper]}\\
    Title: \textcolor{blue}{title}\\
    Abstract:\textcolor{blue}{abstract}\\
    {[The End of Given Paper]}\\
    \\
    {[System]}\\
    Please identify as many relevant aspects from the paper with respect to any research problems in the topic of \textcolor{blue}{folder title}. Once you identified the research problems, describe what specific methods the following paper is applying for each of the problems. Each method from the paper should resolve the matched problem and they should be specific, which means not widely used. Once you identified the methods, describe what specific findings the following paper identified by applying each of the methods.\\ \\
    Finally, return a result as Python dictionary object of the following format: 
    "[\{"Problem": <problem composed of 20-word long phrase>, "Method": <method composed of 20-word long phrase>, "Findings": <findings composed of 20-word long phrase>\}, ..]". If there is no specific method to resolve the problem, then write down "N/A".
}

\subsection{[T3] Generating Paper-paper Descriptions Based on Citances}\label{appendix:desc_citance}

\textbf{System Prompt} 
\texttt{You are an intelligent and precise assistant that can understand the contents of research papers. You are knowledgeable in different fields and domains of science, in particular computer science. You are able to interpret research papers to identify similarities and differences between research papers.}

\noindent\textbf{User Prompt}
\texttt{
    We would like you to compare two research papers for a researcher. You will be provided with the title and abstract of each paper. To help you when you compare the papers, we provided a subsection of Paper A where Paper B is cited. In the subsection of Paper A, cited Paper B already identified methods that are similar between the papers and what problems are solved in each paper using these shared methods.\\
    \\
    {[The Start of Paper A]}\\
    Title: \textcolor{blue}{title}\\
    Abstract:\textcolor{blue}{abstract}\\
    Subsection of Paper A: \textcolor{blue}{subsection of Paper A where Paper B is cited}\\
    {[The End of Paper A]}\\
    \\
    {[The Start of Paper B]}\\
    Title: \textcolor{blue}{title}\\
    Abstract:\textcolor{blue}{abstract}\\
    {[The End of Paper B]}\\
    \\
    {[System]}\\
    Please explain the content of Paper A for a researcher. Explain the paper by comparing it to Paper B, and interpreting the relationships between these papers. Your explanation should only be four sentences long and it should follow the following structure: a sentence that states what aspects are similar between Paper A and Paper B, one sentence summary of Paper A, one sentence summary of Paper B, and one sentence comparing and contrasting between Paper A and B.
}

\subsection{[T4] Paper-paper Descriptions via Generated Pseudo-citances - Finding similar problem aspects across the recommended and collected papers using LLM}\label{appendix:pseudo_similar_aspect}

\textbf{System Prompt} 
\texttt{You are an intelligent and precise assistant that can understand the contents of research papers. You are knowledgeable in different fields and domains of science, in particular computer science. You are able to interpret research papers to identify similarities and differences between research papers.}

\noindent\textbf{User Prompt}
\texttt{
    We would like you to examine a set of papers. You will be given a paper and will be asked to compare this paper to a list of papers labeled A, B, C, and D. You will be provided with the title of each paper and a set of dimensions that describe the content of the paper. These dimensions describe different problems that were addressed by the paper, the method applied in the paper to address each problem, and findings related to that problem and method. These dimensions are provided in a Python JSON format.\\
    \\
    {[The Start of Given Paper]}\\
    Title: \textcolor{blue}{title}\\
    Dimensions:\textcolor{blue}{dimensions}\\
    {[The End of Given Paper]}\\
    \\
    {[The Start of Paper A]}\\
    Title: \textcolor{blue}{title}\\
    Dimensions:\textcolor{blue}{dimensions}\\
    {[The End of Paper A]}\\
    \\
    {[The Start of Paper B]}\\
    Title: \textcolor{blue}{title}\\
    Dimensions:\textcolor{blue}{dimensions}\\
    {[The End of Paper B]}\\
    \\
    {[The Start of Paper C]}\\
    Title: \textcolor{blue}{title}\\
    Dimensions:\textcolor{blue}{dimensions}\\
    {[The End of Paper C]}\\
    \\
    {[The Start of Paper D]}\\
    Title: \textcolor{blue}{title}\\
    Dimensions:\textcolor{blue}{dimensions}\\
    {[The End of Paper D]}\\
    \\
    {[System]}\\
    Please compare the problems of the given paper with the problems of the other listed papers. Please identify papers in the list that have problems that are the most similar with problems in the given paper. Focus on identifying problems that are similar even though they may be resolved with different types of methods. List all of the identified pairs of similar problems, where one problem is from the given paper and the other is a similar problem from another paper in the list.
    For each pair, please describe one shared problem that could contain the two problems. You should avoid just simply concatenating two problems when describing a shared problem. Also, you should avoid containing a phrase that is only included in one of the papers even though it is a very small part.\\ \\
    Finally, return the list of pairs of similar problems and the shared problem for each pair as a list in a Python JSON object of the following format:  
    "[\{"chosen\textunderscore paper": <title of the paper that has a problem that is similar to one in the given paper>, "similar\textunderscore problem": <problem that is similar to a problem in the given paper>, "given\textunderscore problem”: <problem from given paper that is similar to the identified problem>, "shared\textunderscore problem": <one challenge that can encompass the two similar problems>\}, ..]". You should ensure that you return a valid JSON object by escaping any quote marks in your output. (Example: \{"valid\textunderscore object": "This is a "valid" JSON object that escapes any \" characters."\}) If there were no papers that share common problems with the given paper, then only write down "N/A".
}

\subsection{[T5] Paper-paper Descriptions via Generated Pseudo-citances - Verifying whether shared problem is aligned with each paper}\label{appendix:pseudo_verifying}

\textbf{System Prompt} 
\texttt{You are an intelligent and precise assistant that can understand the contents of research papers. You are knowledgeable in different fields and domains of science, in particular computer science. You are able to interpret research papers to identify similarities and differences between research papers.}

\noindent\textbf{User Prompt}
\texttt{
    You will be provided with the title and abstract of Paper A and the given problem.\\
    \\
    {[Title of Paper A]}\\
    \textcolor{blue}{title}\\
    {[The End of the title]}\\
    \\
    {[Abstract of Paper A]}\\
    \textcolor{blue}{abstract}\\
    {[The End of the title]}\\
    \\
    {[The Start of Given Problem]}\\
   \textcolor{blue}{shared problems from paper A and B}\\
    {[The End of Given Problem]}\\
    \\
    {[System]}\\
    Please verify whether Paper A tackled the given problem based on the abstract of the paper. Provide the result as True if Paper A tackled the given problem with their own method, else provide False. If the part of the given problem is not aligned with Paper A's challenges, it should be verified as False. 
}

\subsection{[T6] Paper-paper Descriptions via Generated Pseudo-citances - Generating structured summary}\label{appendix:pseudo_generating_desc}

\textbf{System Prompt} 
\texttt{You are an intelligent and precise assistant that can understand the contents of research papers. You are knowledgeable in different fields and domains of science, in particular computer science. You are able to interpret research papers to identify similarities and differences between research papers.}

\noindent\textbf{User Prompt}
\texttt{
    We would like you to compare two research papers for a researcher. You will be provided with the title of each paper and a set of dimensions that describe the content of the paper. These dimensions describe different problems that were addressed by the paper, the method taken by the paper to address each problem, and findings related to that problem and method. These dimensions are provided in a Python dictionary format. To help you when you compare the papers, we have already identified problems that are similar between the papers and what methods are adopted in each paper to solve the shared problem.\\
    \\
    {[The Start of Paper A]}\\
    Title: \textcolor{blue}{title}\\
    Dimensions:\textcolor{blue}{dimensions}\\
    {[The End of Paper A]}\\
    \\
    {[The Start of Paper B]}\\
    Title: \textcolor{blue}{title}\\
    Dimensions:\textcolor{blue}{dimensions}\\
    {[The End of Paper B]}\\
    \\
    {[The Start of Shared Problems]]}\\
    \textcolor{blue}{shared problem addressed in Paper A and B}\\
    {[The End of Shared Problem]}\\
    \\
    {[The Start of Methods]}\\
    Paper A: \textcolor{blue}{method that is used in Paper A to resolve the aligned problem}\\
    Paper B:\textcolor{blue}{method that is used in Paper B to resolve the aligned problem}\\
    {[The End of Methods]}\\
    \\
     {[The Start of Research interest]}\\
    \textcolor{blue}{folder description}\\
    {[The End of Research Interest]}\\
    \\
    {[System]}\\
    Please explain the content of Paper A for a researcher. Explain the paper by comparing it to Paper B, and interpreting the similarities and differences between these papers. You should consider the researchers’ research interest, which is described above when explaining Paper A. Ensure that your explanation includes information that may be fascinating or engaging for the researcher based on their interests. Your explanation should only be four sentences long and it should follow the following structure: a sentence that states what aspects are similar between Paper A and Paper B, one sentence summary of Paper A, one sentence summary of Paper B, and one sentence comparing and contrasting between Paper A and B.
}

\end{document}